\documentclass[10pt,a4paper,twoside]{article}
\usepackage{epsfig}
\usepackage{baltlat5}
\usepackage{wrapfig}
\newcommand{\PopI}{Pop~I}
\newcommand{\PopII}{Pop~II}
\newcommand{\JKs}{\ensuremath{J\!-\!K_S}}
\newcommand{\JH}{\ensuremath{J\!-\!H}}
\newcommand{\Ha}{H\ensuremath{\alpha}}
\newcommand{\Hb}{H\ensuremath{\beta}}
\newcommand{\Mgb}{Mg~I~b}
\newcommand{\HeI}{He~I}
\newcommand{\NaD}{Na~I~D}
\newcommand{\CaII}{Ca~II}
\newcommand{\Mv}{\ensuremath{M_V}}
\newcommand{\MvsdB}{\ensuremath{M_{V\mathrm{,sdB}}}}
\renewcommand{\Teff}{\ensuremath{T_{\mathrm{eff}}}}
\newcommand{\logg}{\ensuremath{\log g}}
\newcommand{\simgt}{\lower 2pt \hbox{\ensuremath{\, \buildrel {\scriptstyle>}\over{\scriptstyle\sim}\,}}}
\newcommand{\simlt}{\lower 2pt \hbox{\ensuremath{\, \buildrel {\scriptstyle<}\over{\scriptstyle\sim}\,}}}
\newcommand{\sub}[2]{\ensuremath{{#1}_{\mathrm{#2}}}}
\newcommand{\bv}{\ensuremath{B\!-\!V}}
\newcommand{\arcdeg}{\ensuremath{^{\circ}}}

\begin{document}
\ \
\vspace{0.5mm}

\setcounter{page}{1}
\vspace{5mm}

\titlehead{Baltic Astronomy, vol.\ts 14, XXX--XXX, 2005.}

\titleb{THE NATURE OF THE LATE-TYPE COMPANIONS IN HOT SUBDWARF
COMPOSITE-SPECTRUM BINARIES}

\begin{authorl}
\authorb{M.\ A.\ Stark}{1,2} and
\authorb{Richard A.\ Wade}{1}
\end{authorl}

\begin{addressl}
\addressb{1}{Department of Astronomy \& Astrophysics,
Pennsylvania State University, 525 Davey Lab, University Park PA 16802, USA}
\addressb{2}{Current address: Department of Physics \& Astronomy,
University of Wyoming, Dept.\ 3905, 1000 East University Avenue, Laramie WY 82071, USA}
\end{addressl}

\submitb{Received 2005 July ??}

\begin{abstract}
We present the results of a study of the late-type companions in hot
subdwarf composite spectrum binaries.  The exact nature of these
late-type companions has been disputed in the literature --- some
argue that they are main sequence stars, and others have claimed they
are subgiants.  To determine the properties of the late-type
companions, we first conducted a survey utilizing the Two Micron All
Sky Survey (2MASS) All-Sky Data Release Catalog to identify
composite-colored binaries in the {\it Catalogue of Spectroscopically
Identified Hot Subdwarfs} (Kilkenny, Heber, \& Drilling 1988, 1992).
We then conducted a spectroscopic study of a sub-sample of the 2MASS
composite-colored hot subdwarfs.  The sample consists of
photometrically and spectroscopically single and composite hot
subdwarfs (14 single and 51 composite).  We also obtained spectra of
59 single late-type stars with Hipparcos parallaxes for calibration.
We used measured equivalent width (EW) indices from the composite
systems to estimate the temperature and gravity of the late-type star,
taking into account the dilution of its spectral features by light
from the hot subdwarf.  Results from combining the spectroscopic data
with model energy distributions indicate that the late-type companions
in composite-spectrum systems are best described by main sequence
companions overall.
\end{abstract}

\vskip4mm


\resthead{\LaTeX\ style for Baltic Astronomy}{M.~A.~Stark and R.~A.~Wade}

\sectionb{1}{DEFINING THE SAMPLE}

For this investigation we studied hot subdwarf stars listed in the
{\it Catalogue of Spectroscopically Identified Hot Subdwarfs}
(Kilkenny, Heber, {\&} Drilling 1988, KHD) as updated and expanded in
an electronic version by D.~Kilkenny, c.\,1992.\@ While the KHD
catalog contains all varieties of hot subdwarfs, we primarily focused
on the more numerous sdB stars. The sdB are understood to be
relatively homogeneous and probably have a common evolution history
from the zero-age extended horizontal branch (ZAEHB), while sdO stars
likely follow multiple evolutionary pathways and might be expected to
be less homogeneous and to have less simply explained properties.

To make a comparison of the KHD data with existing databases (such as
2MASS) or to obtain new observations of the correct star, accurate
coordinates on a consistent system are required.  For each entry the
object's position was verified by referring whenever possible to
original published finding charts or by contacting knowledgeable
observers, then locating the object on a chart prepared from the USNO
A2.0\ Catalog (Monet 1998, see also Stark {\&} Wade 2003).\@

\sectionb{2}{2MASS RESULTS}

We collected readily available visible and near-IR flux measurements
of hot subdwarfs from the 2MASS All-Sky Data Release (ASDR) Catalog
and identified those whose colors indicate the presence of a late type
companion (for more information see Stark {\&} Wade 2003; Stark, Wade,
{\&} Berriman 2005).\@ We thus determined the fraction of hot
subdwarfs that exist in composite spectrum binaries ($\sim$40{\%} of
sdBs from KHD are composite in a magnitude limited sample).  We
defined an approximately volume limited sample of hot subdwarfs from
KHD for statistical purposes (see Stark {\&} Wade 2003; Stark, Wade,
{\&} Berriman 2005), and found that $\sim$25{\%} of sdBs are composite
in a volume limited sample (VLS).

We defined the color parameter $Q = 0.752 (\JH) + (\JKs)$, which gives
the clearest separation between composite and single hot subdwarfs
based on 2MASS photometry alone.  We compared the distributions in
{\JKs}, {\JH}, and $Q$, and found them all to show a bimodally
distributed population (Figure~1).  In a histogram of the IR color
indices {\JKs} and $Q$, the two peaks of the bimodal distribution can
be understood as single stars (blue peak at $\JKs = -0.170$,
$Q\approx-0.275$) and composite systems (red peak at $\JKs = +0.289$,
$Q\approx+0.500$).  This bimodal distribution is also present in the
approximately VLS, again with the two peaks at $\JKs = -0.167$ and
+0.248, and $Q \approx -0.275$ and +0.475.

{\it There are no (or very few) F or dM companions of the hot
subdwarfs in the KHD catalog}.  This is evident from the bimodal
distribution in 2MASS colors ($Q$, {\JKs}, and {\JH}).  Were there a
large population of F or dM companions, their composite colors would
have filled in the gap between the two bimodal peaks.  However, the
distribution in 2MASS colors can be described by only a very small (or
no) spread in the colors of the late-type companions.  In the case of
F-type and earlier companions, should they actually exist, it is
likely that most of them were never identified as containing a hot
subdwarf.  The F-type star would dominate the light at visual
wavelengths, and the combined light would look spectroscopically like
a metal-poor {\PopII} star (the metal lines of the {\PopI} star being
diluted by the hot subdwarf so they look like a {\PopII} star).  So,
it is understandable that there are very few of these objects in the
current KHD catalog.  The dM stars on the other hand, have no obvious
reason to be selected against in surveys that have identified hot
subdwarfs.  The dM is significantly fainter than the hot subdwarf, so
that it should be basically undiscernible in the visible (both
photometrically and spectroscopically).  So, the fact that there are
no (or very few) dM companions in the KHD sample represents a true
trend in the hot subdwarf population (as opposed to a possible
selection bias as in the case of the F-type and earlier stars).

The observed distribution of hot subdwarfs in 2MASS colors can be
reproduced equally well by either assuming main sequence companions
with $\MvsdB \approx 4.5-5.0$~mag, or by assuming subgiant companions
with more luminous sdB stars ($\MvsdB \approx 2.5-3.0$~mag) ---
photometric data alone cannot distinguish between these two
possibilities.

\sectionb{3}{SPECTROSCOPY OF COMPOSITE HOT SUBDWARFS}

Spectroscopy of a sub-sample of the 2MASS composite-colored hot
subdwarfs was obtained to break the degeneracy between main sequence
and subgiant companions present in the 2MASS and visual photometry
alone.  Observations were made primarily at the Kitt Peak National
Observatory (KPNO) 2.1m\ telescope using the GoldCam spectrograph, but
some additional observations came from the McDonald Observatory 2.7m\
telescope with LCS.\@ Both sets of observations cover roughly
4500--9000~{\AA} with $\sim$3.3~{\AA} resolution ($\sim$1.3~{\AA}/pix)
using two spectrograph settings.\@ This wavelength region covers
{\Hb}, {\Mgb}, {\HeI} 5875~{\AA}, {\NaD}, {\Ha}, {\HeI} 6678~{\AA},
and the {\CaII} IR Triplet (CaT).  The sample of observed stars
consists of photometrically and spectroscopically single and composite
hot subdwarfs (14 single and 51 composite).  We also obtained spectra
of 59 single late-type stars from both the main sequence and subgiant
branch with Hipparcos (HIP) parallaxes for calibration.  Example
spectra from KPNO GoldCam (of a single, composite, and standard star)
are shown in Figure~2.  Our analysis focused on {\Mgb}, {\NaD}, and
CaT equivalent widths (EWs).  Each of these lines has a very different
behavior with {\Teff} and {\logg},
so they are useful for constraining the dilution by the hot subdwarf,
as well as {\Teff} and {\Mv} of the late-type companion, thus breaking
the main sequence-subgiant degeneracy present in the 2MASS and visual
photometry alone.\@

The observations (2MASS and visual photometry combined with EWs) for
each composite hot subdwarf were compared with diluted models based on
HIP standard star observations, models of ZAEHB stars (Caloi 1972),
terminal-age EHB (TAEHB) stars (Dorman, Rood, {\&} O'Connell 1993),
and Kurucz (1998) spectral energy distributions, in order to determine
the combination of sdB+late-type star that best explained all
observations.  In most cases the actual fit was driven primarily by
the measured EWs, and secondarily by {\JKs} color (this agrees with
the previous determination that photometry alone cannot distinguish
between main sequence and subgiant companions in these cases).  With a
few exceptions, it was found that the late-type companions in
composite-spectrum systems are best identified as main sequence.  The
majority of the well constrained main sequence companions have $0.5
\simlt \sub{(\bv)}{comp} \simlt 1.1$ (spectral types $\sim$F6--K5, see
Figure~3).  The spectra and identifications of four composite
subdwarfs are compared in Figures~4 and 5.

There are some interesting objects identified through our spectroscopy.
These include:
\begin{itemize}
\setlength{\topsep}{0pt}
\setlength{\parsep}{0pt}
\setlength{\itemsep}{0pt}
 \item Two new emission line objects, 
 LS~IV$-08\arcdeg03$ (possible x-ray binary) and
 PB~5333 (NLTE emission in the core of {\Ha}). One possible new NLTE
 core emission object, TON~264.

 \item Nine objects that are best fit with subgiant companions
 (assuming ZAEHB or TAEHB hot subdwarfs), with an additional six best
 fit with subgiants assuming TAEHB hot subdwarfs.  PG~0232+095's
 late-type companion seems to show molecular features indicative of a
 giant star (e.g.,\ possible CN-red molecular bands), but the CaT
 appears too weak for a giant star --- this object requires further
 study.

 \item A possible {\it resolved}\/ visual double sdB+sdB (or sdB+HBB), 
 HZ~18 (Figure~6), which may also contain an inner short-period binary
 (based on the velocity difference between the spectra for the two
 stars).
\end{itemize}

There were 18 cases in which the late-type companion was poorly fitted
by our models (namely the best-fit parameters fell right at or near
the edge of our model grid).  The EWs in some of these objects may be
erroneous due to an interstellar contribution.  Additional refinement
of the models, extension of the models to include a larger temperature
range in both the late-type stars (by obtaining more observations of
standards) and the hot subdwarfs (more models over a larger
temperature range), or adjustments to correct for interstellar
contributions, are needed to accurately fit these objects.

\sectionb{4}{LIMITATIONS AND DIRECTIONS FOR FUTURE WORK}

Our modelling procedure is limited in the range of both hot subdwarf
and late-type stars included.  These models could be greatly improved
by including a greater range in temperatures for both the companion
and particularly for the hot subdwarf.  Specifically we have trouble
identifying and coping with the hottest sdBs, and the sdOs.  We are
also using the assumption that the hot member is in fact a true
sdB-type star; if it is in fact a HBB or post-EHB star, then the
modelling breaks down, giving bogus fits.  Additional information to
help constrain the properties of hot subdwarf would be of value
(including whether it is sdB, sdO, post-EHB, or HBB).  This additional
information could include UV observations or spectra with coverage
farther to the blue.  Also, including additional late-type spectral
features from our spectra would help better constrain the fits.

Future work related to, or stemming from, this project includes:
\begin{enumerate}
\setlength{\topsep}{0pt}
\setlength{\parsep}{0pt}
\setlength{\itemsep}{0pt}
\item Classification of more composite hot subdwarfs.

 \item Long-term RV studies of composite spectrum systems to determine
 periods (or at least set lower limits on the periods).

 \item Follow-up observations of ``unusual'' objects identified,
 including (for example): PG~0232+095, TON~264, 
 HZ~18, and the emission lined objects. 

 \item Further observations of the resolved visual doubles, including
 proper motions, and better classifications of the companions
 (particularly HZ~18, which may be a resolved sdB+sdB or sdB+HBB system).
\end{enumerate}

\sectionb{5}{IMPLICATIONS} \label{ch:summary:Implications}

Han et al.\ (2002, 2003) predict that for companions that are later
than $\sim$G, all companions in short-period systems are main sequence
stars (in post-CE binaries) and all companions in long-period systems
(P $>$ 40 days) are subgiant or giant stars (in post-Roche lobe overflow
binaries).  RV studies of composite spectrum hot subdwarfs with
FGK-type companions (i.e.,\ Orosz, Wade, \& Harlow 1997; Maxted et
al.\ 2001; Saffer, Green, \& Bowers 2001), have found that the orbital
periods must be long, many months to years or more.  In the Han et
al.\ scenario this would imply that they contain subgiant or giant
companions. Han et al.,\ however, assume that hot subdwarfs with
subgiant and giant companions, i.e.,\ these same long-period systems,
were excluded from surveys for hot subdwarfs.  Indeed, in our study,
the majority of composite companions are consistent with main sequence
stars (although we have identified some subgiant companions, so this
exclusion is not complete).

If the GK-type companions are main sequence stars, why do they seem
to be in long-period binaries?  At face value, there is something
incorrect or incomplete in the Han et al.\ binary formation scenario or
its interpretation as applied to existing samples of hot subdwarfs.
It may be that aspects of the Han et al.\ study (binary evolution model, or
mapping onto observables) are at the heart of the matter; or perhaps
the apparent contradictions can be resolved via discovering some
subtlety of different sample selection for the RV studies reported so
far and our present spectroscopic analysis. (The latter possibility
can be assessed, for example, by an RV study of the composite binaries
in our study.)

\vskip4mm
\vbox{
\centerline{\psfig{figure=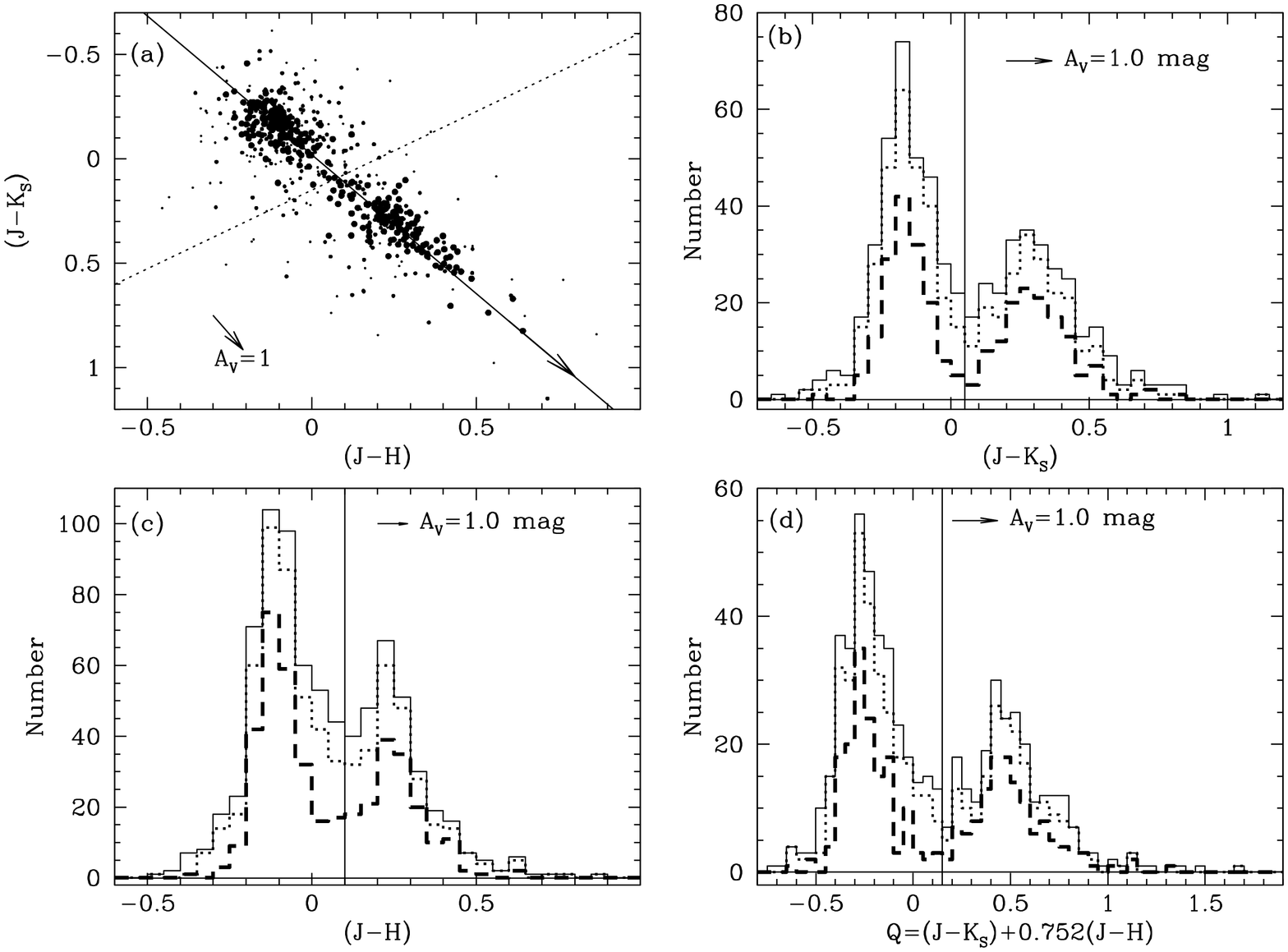,width=84truemm,angle=-90,clip=}}
\captionc{1}{Panel $a$: Color-Color
  plot for 2MASS colors of sdBs only with a linear fit to the points
  shown as the solid line, the parameter $Q$ increases along this line
  as indicated.  The dashed line is mathematically perpendicular to
  the linear fit line, and demonstrates a contour of constant $Q$ (the
  two lines cross at $Q=+0.15$).  All other panels contain histograms
  (bin sizes 0.05) of 2MASS color indices: {\JKs} (panel $b$), {\JH}
  (panel $c$), and $Q$ (panel $d$).  In the three histograms (panels
  $b$--$d$), the solid line is for all sdB, the dotted line is for sdB
  with both $\sigma(\JKs)$ and $\sigma(\JH)<0.2$ the dashed line is
  for sdB with both $\sigma(\JKs)$ and $\sigma(\JH)<0.1$.  Single sdBs
  fall in the left peak ($Q<0.15$), composites fall in the right peak
  ($Q\geq0.15$).  The effect of 1 magnitude of extinction ($A_V = 1$)
  is indicated.}
}
\vskip4mm

\vbox{
\centerline{\psfig{figure=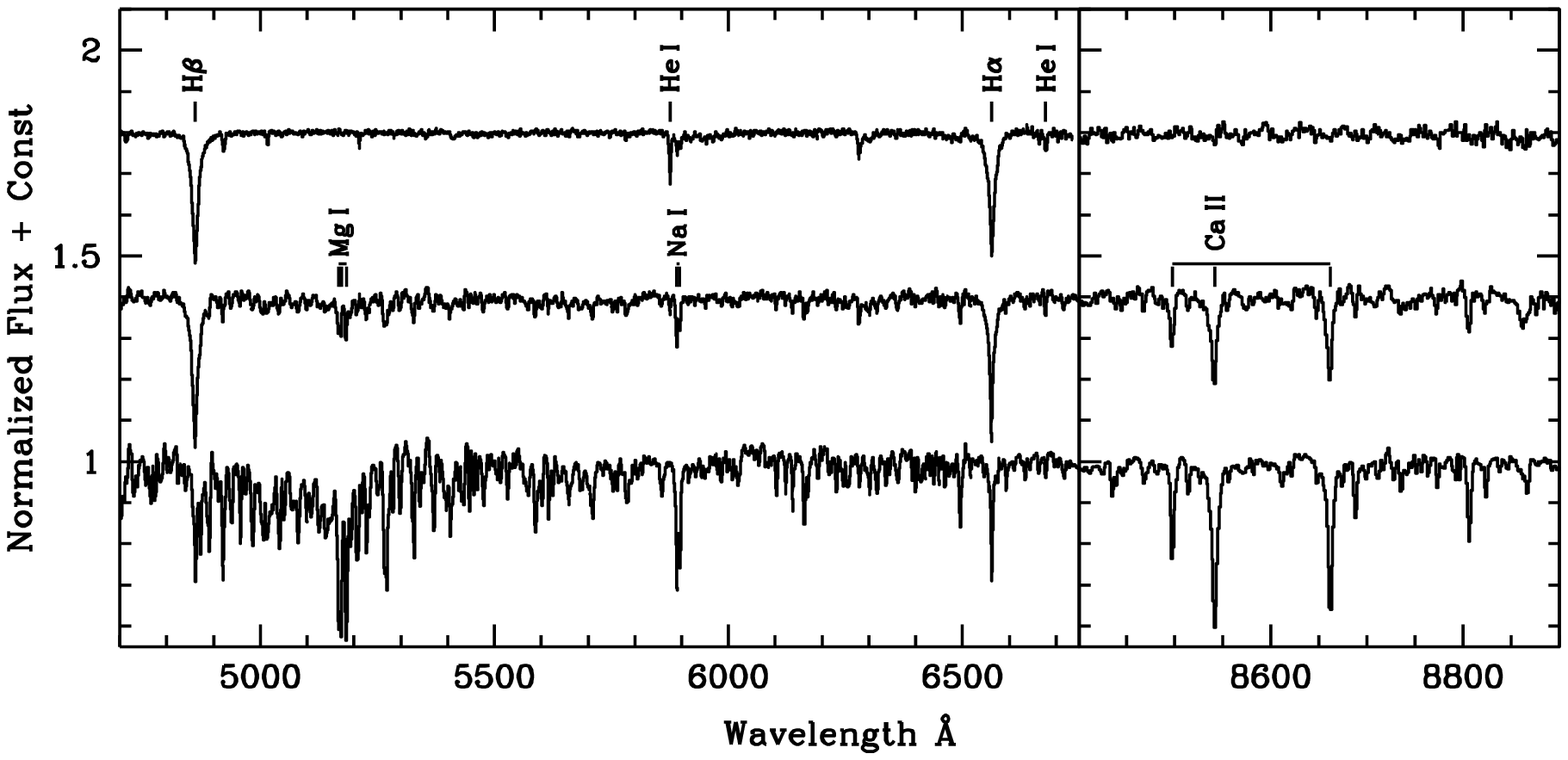,width=109truemm,angle=0,clip=}}
\captionc{2}{Normalized spectra for a ``typical'' single hot subdwarf (top, 
  LS~IV$+00\arcdeg$21), a composite hot subdwarf (middle, 
  PB~6107),  and a single HIP standard star (bottom, HIP~13081).  Left 
  panel shows the region from {\Hb} to {\Ha}, right panel shows the
  CaT.  Prominent spectral features are labelled.}  
}
\vskip4mm

\vbox{
\centerline{\psfig{figure=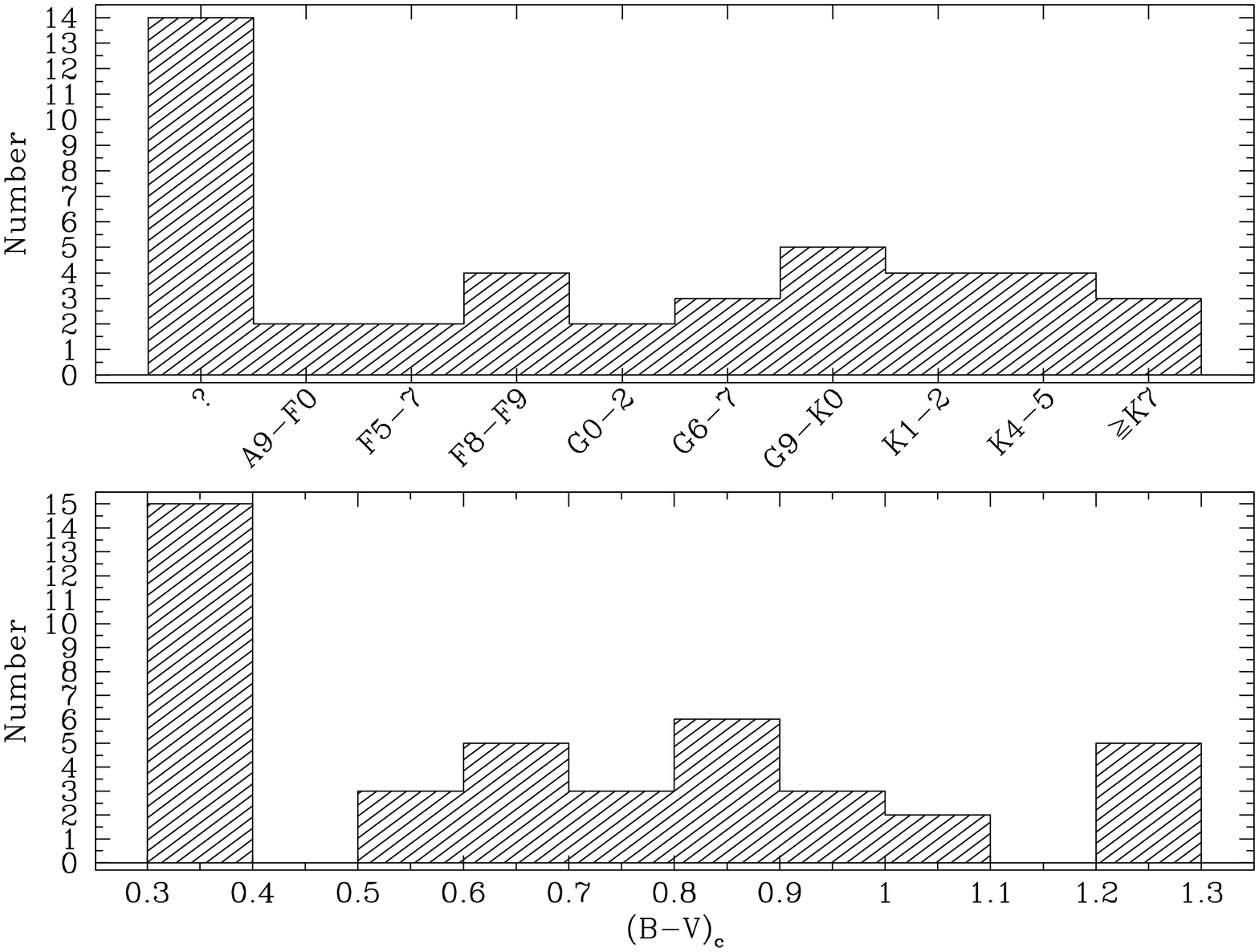,width=84truemm,angle=-90,clip=}}
\captionc{3}{Histogram showing the distribution of late-type companions
  in {\bv} (bottom) and approximate Spectral Type (top).  Objects in
  the last bins on either end of the histograms are upper or lower
  limits and either (1)~have problems with their fits or (2)~belong
  outside the range of parameters examined (see discussion at the end of \S3).
  PG~0232+095, and those stars fitted with subgiant companions (assuming
  ZAEHB hot subdwarfs)
  have been excluded from this plot.}
}
\vskip4mm

\vbox{
\centerline{\psfig{figure=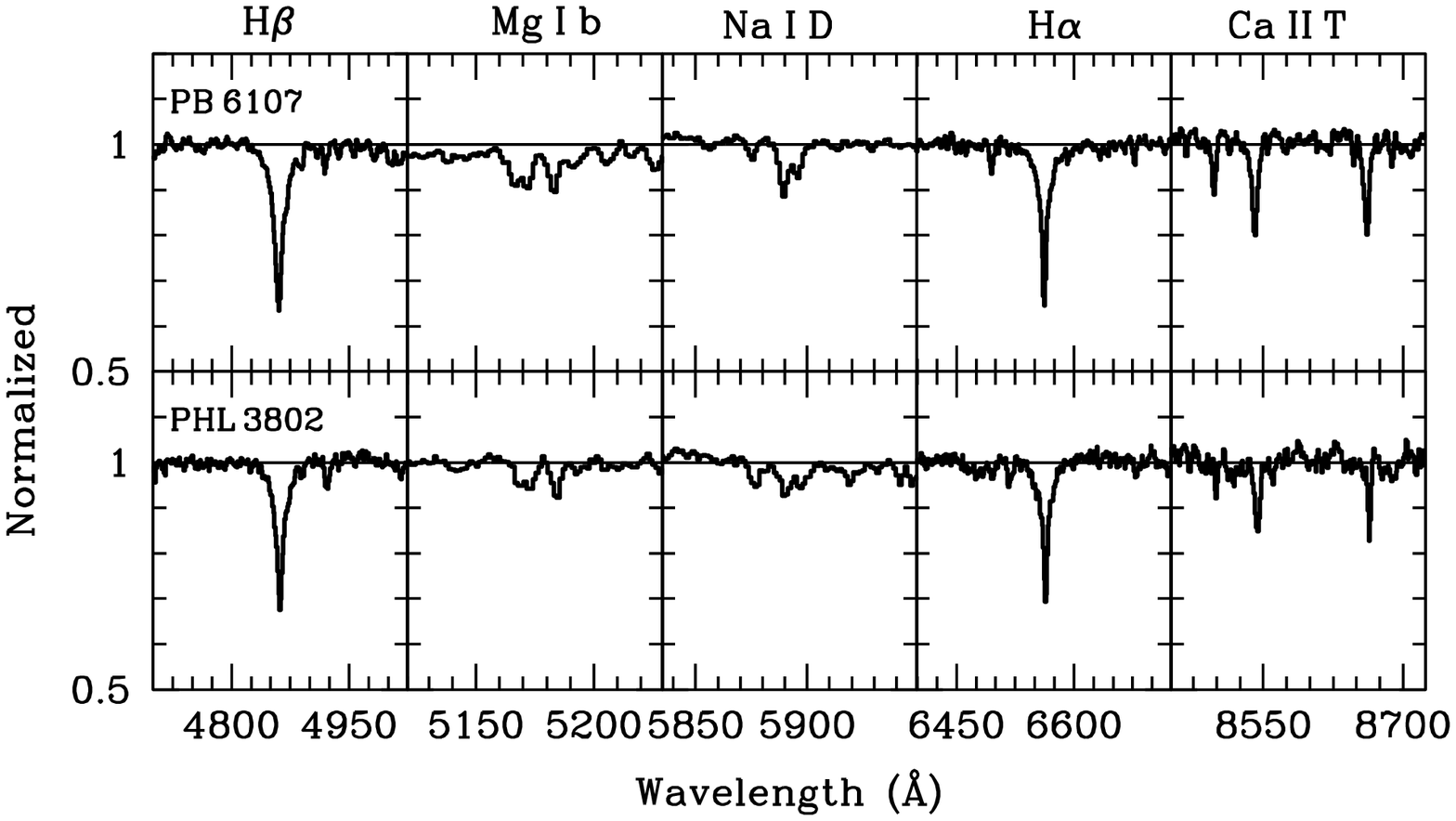,width=109truemm,angle=0,clip=}}
\captionc{4}{Comparison of the major spectral features in the spectra 
  of PB~6107 and PHL~3802.  Panels from left to right show: {\Hb},
  {\Mgb}, {\NaD}, {\Ha}, and CaT.  These two stars were both fit with
  similar hot subdwarfs (${\bv}=-0.246$ {\&} $-0.233$ and ${\Mv}=4.43$
  {\&} 4.22 for PB~6107 and PHL~3802 respectively) and companions
  (G9V--K0V), so their spectra look similar.  (Continuum fits shown
  were {\it not} used for the calculation of EWs.)}
}
\vskip4mm

\vbox{
\centerline{\psfig{figure=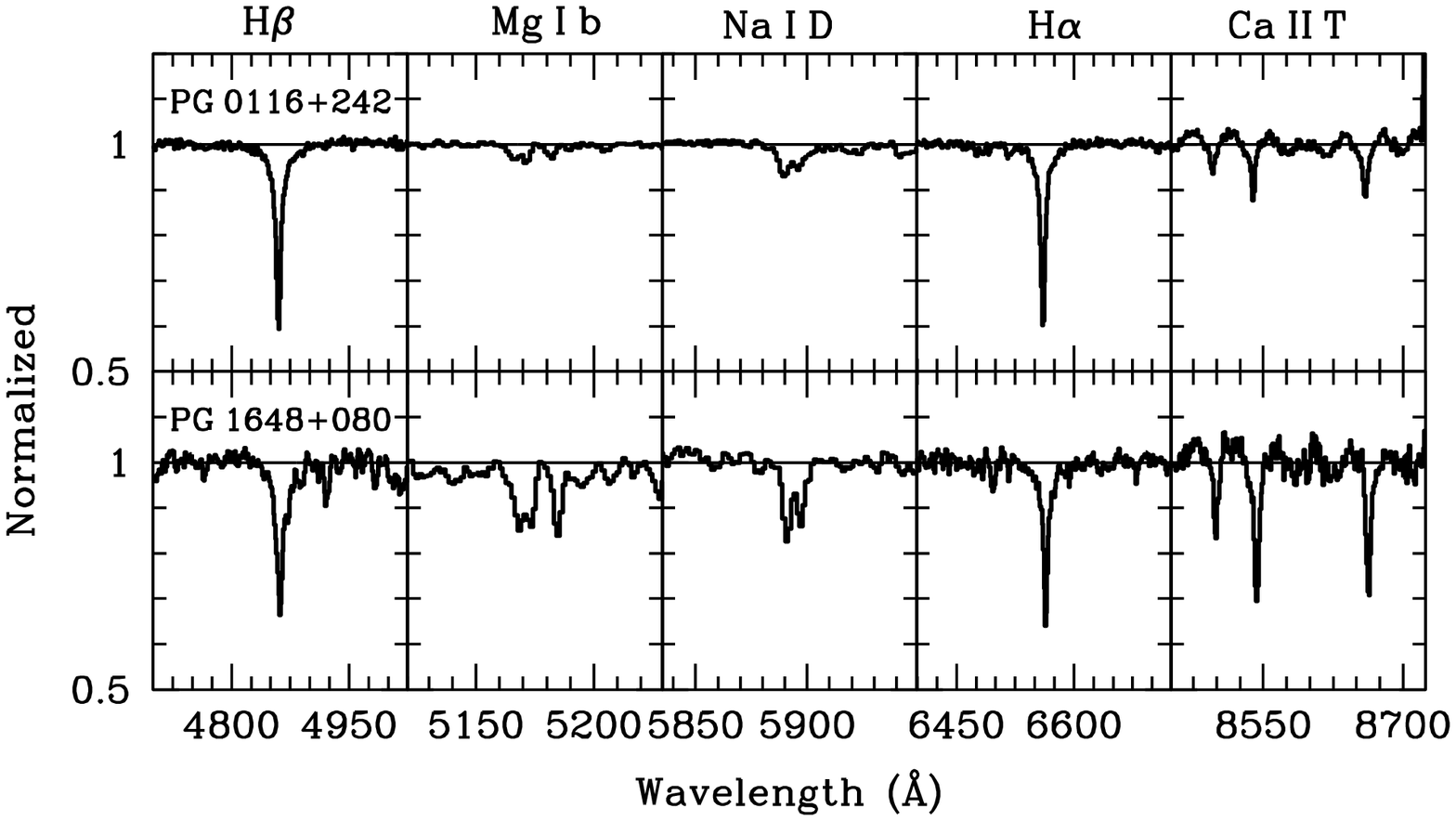,width=109truemm,angle=0,clip=}}
\captionc{5}{Comparison of the major spectral features in the spectra 
  of PG~0116+242 and PG~1648+080.  Panels from left to right show:
  {\Hb}, {\Mgb}, {\NaD}, {\Ha}, and CaT.  These two stars were fit
  with similar companions (G0IV) but different hot subdwarfs
  (PG~1648+080 was fit with a hotter, fainter hot subdwarf, while
  PG~0116+242 was fit with a cooler, brighter hot subdwarf).  In
  PG~1648+080, the late-type companion dominates over the hot subdwarf
  so its lines appear much stronger in the combined spectra, while in
  PG~0116+242, the brighter hot subdwarf washes out the features from
  the late-type star making them appear much weaker.  (Continuum fits
  shown were {\it not} used for the calculation of EWs.)}
}
\vskip4mm

\vbox{
\centerline{\psfig{figure=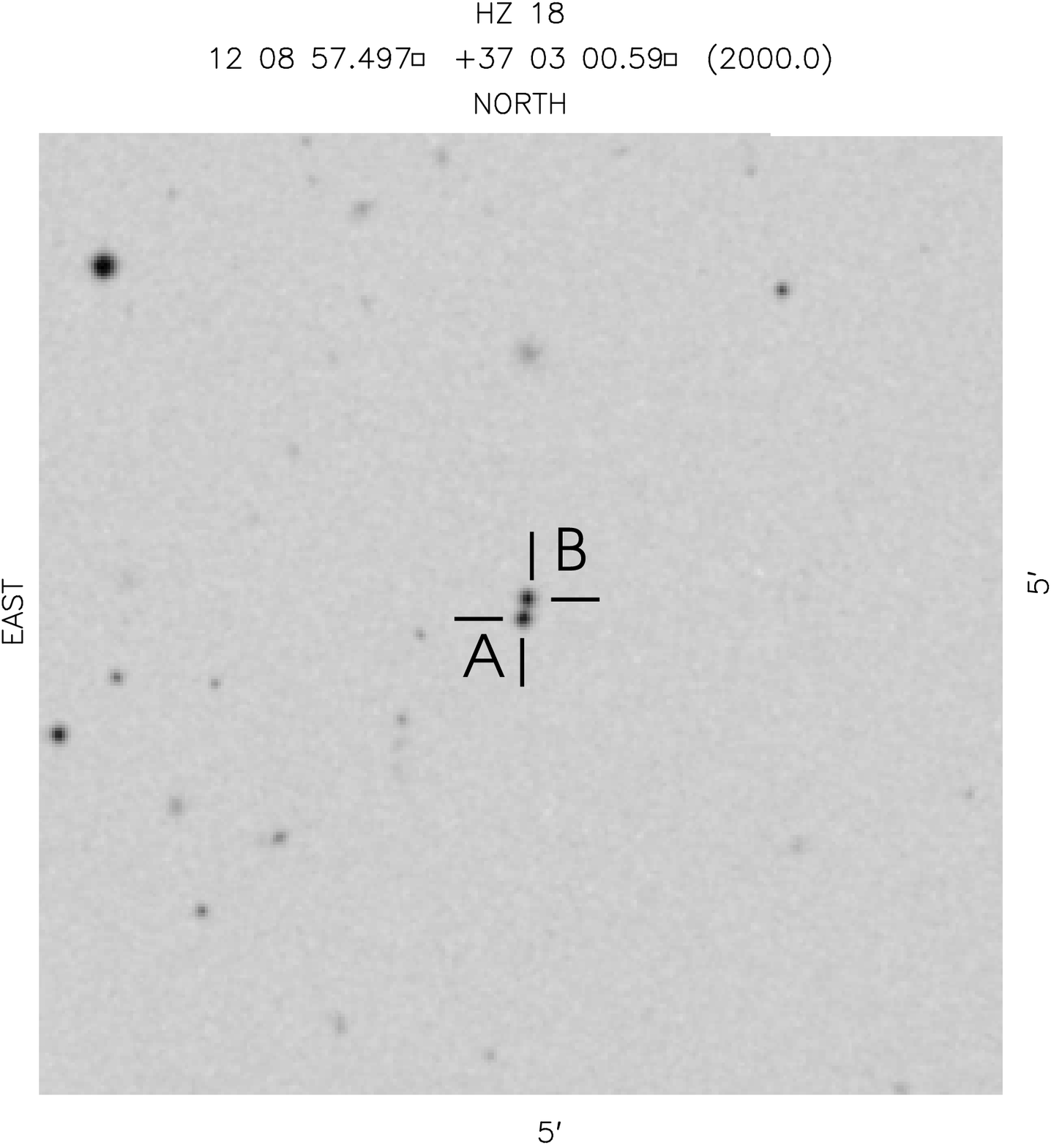,width=42truemm,angle=-90,clip=}\psfig{figure=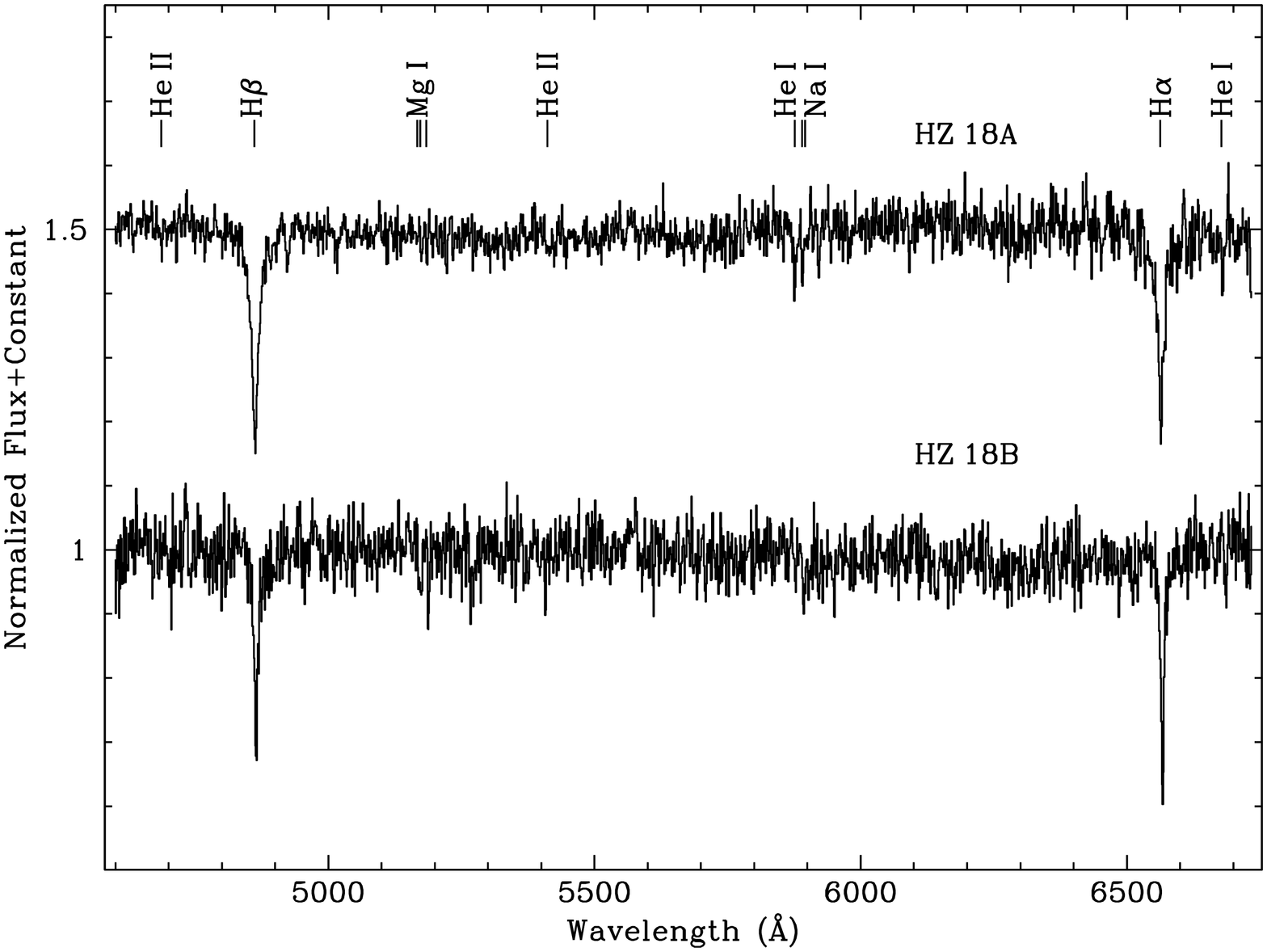,width=42truemm,angle=-90,clip=}}
\captionc{6}{Finding chart and spectra for the two components of HZ\,18 (=A).}
}
\vskip4mm

ACKNOWLEDGMENTS.  This research has been supported in part by:
NASA grant NAG5--9586, NASA GSRP grant NGT5--50399,
Zaccheus Daniel Fund for astronomy research,
Sigma Xi Grant-in-Aid of Research,
NOAO Thesis Travel Support, and
a NASA Space Grant Fellowship through the Pennsylvania Space Grant
  Consortium.
This research has made use of: data products from the Two Micron
All Sky Survey (2MASS), the SIMBAD database, the Digitized Sky Surveys,
and the NASA/IPAC Extragalactic Database (NED).

\goodbreak

\References

\refb
{Caloi}, V. 1972, A\&A, 20, 357

\refb
{Dorman}, B., {Rood}, R.~T., \& {O'Connell}, R.~W. 1993, ApJ, 419, 596

\refb
{Han}, Z., {Podsiadlowski}, P., {Maxted}, P.~F.~L., \& {Marsh}, T.~R. 2003,
  MNRAS, 341, 669

\refb
{Han}, Z., {Podsiadlowski}, P., {Maxted}, P.~F.~L., {Marsh}, T.~R., \&
  {Ivanova}, N. 2002, MNRAS, 336, 449

\refb
{Kilkenny}, D., {Heber}, U., \& {Drilling}, J.~S. 1988, South African
  Astronomical Observatory Circular, 12, 1

\refb
{Kurucz}, R. 1998, {\it Solar abundance model atmospheres for 0,1,2,4,8 km/s.}
  (Cambridge, Mass.:\ Smithsonian Astrophysical Observatory), \\
  http://kurucz.harvard.edu/

\refb
{Maxted}, P.~F.~L., {Heber}, U., {Marsh}, T.~R., \& {North}, R.~C. 2001,
  MNRAS, 326, 1391

\refb
{Monet}, D.~B.~A., {Canzian}, B., {Dahn}, C., {Guetter}, H., {Harris}, H.,
  {Henden}, A., {Levine}, S., {Luginbuhl}, C., {Monet}, A.~K.~B., {Rhodes}, A.,
  {Riepe}, B., {Sell}, S., {Stone}, R., {Vrba}, F., \& {Walker}, R. 1998, {The
  USNO-A2.0 Catalogue} (U.S. Naval Observatory Flagstaff Station [USNOFS] and
  Universities Space Research Association stationed at USNOFS)

\refb
{Orosz}, J., {Wade}, R.~A., \& {Harlow}, J.~J.~B. 1997, AJ, 114, 317

\refb
{Saffer}, R.~A., {Green}, E.~M., \& {Bowers}, T. 2001, in ASP Conf.\ Ser.\ 226:
  12th European Workshop on White Dwarfs, 408

\refb
{Stark}, M.~A. \& {Wade}, R.~A. 2003, AJ, 126, 1455

\refb
{Stark}, M.~A., {Wade}, R.~A., {\&} Berriman, G.~B. 2004, Astrophysics and
  Space Science, 291, 333

\end{document}